\documentclass[
reprint,
superscriptaddress,
aps,
prl,
]{revtex4-2}
\usepackage{graphicx}
\usepackage{bm}
\usepackage[usenames,dvipsnames]{color} %using the color package, not xcolor

\usepackage{multirow}
\usepackage{xr-hyper}
\usepackage{hyperref}
\usepackage{subfigure}
\usepackage{physics}
\usepackage[usenames,dvipsnames]{color}
\usepackage{dcolumn}% Align table columns 
\begin{document}

%\title{Phase-corrected terahertz reflection spectroscopy reveals near-zero-index behaviour in indium antimonide}
\title{Near-zero-index behavior in indium antimonide revealed by phase-corrected terahertz reflection spectroscopy}

\author{Kasturie D. Jatkar}
\affiliation{Department of Physics, Stockholm University, 106 91 Stockholm, Sweden.}
\author{Tien-Tien Yeh}
\affiliation{Department of Physics, Stockholm University, 106 91 Stockholm, Sweden.}
\author{Matteo Pancaldi}
\affiliation{Department of Molecular Sciences and Nanosystems, Ca’ Foscari University of Venice, 30172 Venice, Italy}
\author{Stefano Bonetti}
\email{stefano.bonetti@fysik.su.se}
\affiliation{Department of Physics, Stockholm University, 106 91 Stockholm, Sweden.}
\affiliation{Department of Molecular Sciences and Nanosystems, Ca’ Foscari University of Venice, 30172 Venice, Italy}

% PRL word limit: 3750 words
%Displayed Math	The word equivalent for displayed math is 16 words per row for single-column equations. Two-column equations count as 32 words per row. (16*12 = 192)
%Figures	To estimate the word equivalent for figures use the figure’s aspect ratio (width / height). The estimate is [(150 / aspect ratio) + 20 words] for single-column figures, and [300 / (0.5 * aspect ratio)] + 40 words for double-column figures.

\begin{abstract}
%Terahertz time-domain spectroscopy is a well established technique used for measuring optical properties of materials. However, its applicability has been limited to transparent materials, mainly due to its strict requirement for high precision in the placement of the sample and reference. This leads to large errors in the results extracted from a reflection measurement, which must be corrected using highly iterative codes or with heavy post-processing.
We developed a phase correction method for broadband terahertz time-domain spectroscopy in reflection geometry, which allows us to obtain quantitative and accurate values for the complex refractive index of materials. The process is analytical, based on the Kramers-Kronig relations, and does not require any computationally intensive algorithms. We validate it by extracting the refractive index of silicon, obtaining the nominal value with an accuracy better than 2.5\% over the 0.25--3.5 THz range, and better than 0.6\% in the 1--2 THz range. We use the method to experimentally observe that an undoped InSb crystal shows a refractive index of $n<1$ between 1 and 2 THz, in proximity to the plasma frequency of the material where, the amplitude of the group velocity goes as low as 0.08$c$.
\end{abstract}

\maketitle

%\tableofcontents

%\ intro
The technology used in terahertz (THz) time-domain spectroscopy (TDS) has rapidly developed over the past few decades. THz-TDS has proven to be an extremely useful tool for research in physics \cite{hirashita2016firstgeneration, dhillon20172017, spies2020terahertz, seo2022terahertz}, chemistry \cite{baxter2011terahertz}, biology \cite{son2009terahertz, nikitkina2021terahertz}, and it is also of interest for applications in high-speed communications \cite{kim2017910m, harter2020generalized}, security \cite{kemp2003security, palka2012THz}, and food technology \cite{afsah-hejri2019comprehensive, feng2021terahertz}.
In the field of condensed matter physics, terahertz photons, with a characteristic energy of few meV, are particularly suitable to study low-energy collective excitations such as phonons \cite{salen2019matter}, polarons \cite{vanmechelen2008electronphonon}, polaritons \cite{kojima2003farinfrared,wu2022manipulating}, and magnons \cite{neeraj2021inertial, unikandanunni2021anisotropic, qin2015longliving, mashkovich2021terahertz}. Even the quest for the axion particle is exploring the THz range for its detection \cite{dona2022design,schutte-engel2021axion}, making THz-TDS, with its intrinsic broadband nature, a valuable approach to test potential detectors over a wide frequency range.
%In condensed matter physics, the meV range photon energy proves to be beneficial for observation of quasi-particles and collective excitations, and invokes plenty of cutting-edge research, such as axions\cite{axion_Huang_Yuan_2016,axion_Barry_2021}, polarons\cite{polaron_vanMechelen_2008,polaron_Zheng_2021}, polaritons\cite{polariton_Kojima_2003,polariton_Wu_2022}, spintronics\cite{Neeraj_2021,spin_Turenne_2022}, etc.

A distinguishing characteristic of THz-TDS is that the electric field detection provides not only the amplitude of the signal but also its phase. This allows, in principle, to extract the complex optical properties of materials \cite{withayachumnankul2009engineering, jepsen2019phase}. However, such a procedure is in practice very complicated, especially when performed in reflection geometry, where a precise alignment between the sample and a high-reflective reference (e.g., a gold mirror) is essential. This geometry, though, is the one of greatest interest and applicability, given that only a fraction of materials transmit a substantial amount of THz radiation over a broad frequency range. Remarkably, a misplacement of the order of 1 \textmu m between the sample and the reference, which would appear negligible compared to the THz wavelength (1 THz corresponds to 300 \textmu m), creates enormous artifacts in the extraction of the optical properties \cite{pashkin2003phasesensitive}. Hence, despite its significance, only a limited number of studies have been conducted in reflection geometry due to the demanding sub-micrometer alignment requirements. To minimize alignment errors and retrieve the phase accurately, a few strategies have been implemented: samples have been measured using self-referencing techniques such as ellipsometry \cite{neshat2012terahertz} or by combining results from measurements in transmission and reflection geometry \cite{mori2014progress}.
%, and pump-probe spectroscopy \cite{Hempel_Unold_Eichberger_2017,Pizzuto_2021}. The use of a femtosecond-pulsed laser or a half coated-sample/reference for calibrating the experiment has also been reported \cite{Lee_2014}, which requires additional fabrication steps.
%Nowadays, except for self-referencing measurements such as ellipsometry\cite{Neshat_Armitage_2012}, combined measurements in transmission and reflection \cite{STO_tr_Mori_Igawa_Kojima_2014}, and pump-probe spectroscopy\cite{Hempel_Unold_Eichberger_2017,Pizzuto_2021}, the fs-pulsed laser and half coated-sample/reference have been experimentally applied for calibration\cite{Pashkin_2003,Lee_2014}(?????). 
An alternative approach, explored in literature, deals with the alignment challenge as a mathematical problem solvable using various techniques. Some of these include the maximum entropy model \cite{vartiainen2004numerical, vartiainen2013optical, gornov2006comparison}, subtractive Kramers-Kronig relations (SKK) based methods such as singly-SKK \cite{lucarini2005detection,gornov2006comparison}, multiply-SKK \cite{palmer1998multiply}, and differential multiply-SKK \cite{granot2008differential}. While effective within certain constraints, these methods translate an experimental complexity into a numerical one, with computationally heavy algorithms that seem to be not necessary for a problem that can be simulated analytically.

In this Letter, we introduce a technique employing the Kramers-Kronig relations to find the precise experimental misalignment of the sample relative to the reference. This method exhibits exceptional accuracy at the sub-micrometer scale. Consequently, we can effectively correct the frequency-dependent phase in THz-TDS measurements within the frequency range of 0.25 to 4 THz of our setup. We demonstrate the applicability of our phase correction method by extracting the complex refractive index of silicon and indium antimonide crystals. We describe the essential mathematical procedures and experimental validation in the context of reflection geometry at normal incidence. Further mathematical details and the generalization towards arbitrary incidence angle and polarization are given in the companion paper \cite{UDCM}.

\begin{figure}[t!]
    \centering
    \includegraphics[width=\columnwidth]{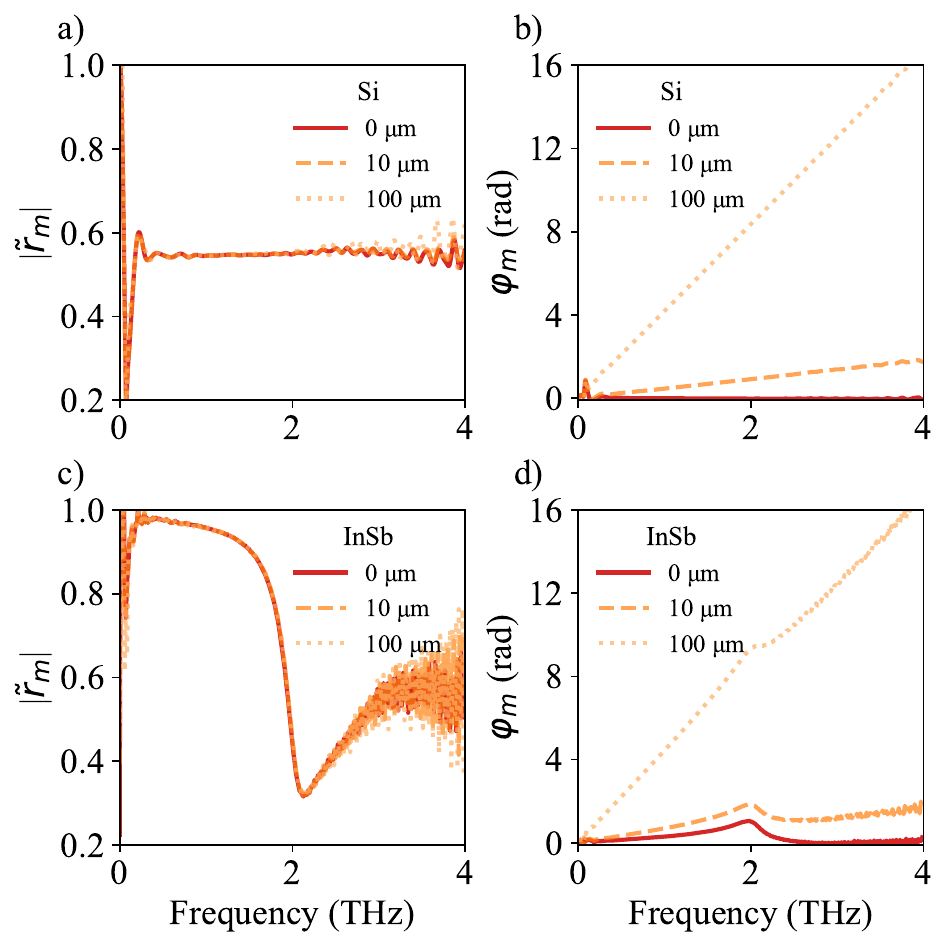}
    \caption{(a) Magnitude $|\tilde{r}_{m}|$ and (b) phase $\varphi_m$ of the reflectivity measured in Si at three relative shifts: 0, 10, and 100 \textmu m. (c) Magnitude and (d) phase of InSb at the same relative shifts.}
    \label{fig:fig1}
\end{figure}
%\begin{figure}[h]
 %    \centering
 %    \begin{subfigure}
 %        \centering         
 %        \includegraphics[width=0.224\textwidth]{Fig1/Si_lnr_all.pdf}
       
     %\end{subfigure}
     %\begin{subfigure}
      %   \centering
      %   \includegraphics[width=0.22\textwidth]{Fig1/Si_phi_all.pdf}
         %\caption{(b)}
     %\end{subfigure}
     %\centering
     %\hfill
     %\begin{subfigure}
     %    \centering
     %    \includegraphics[width=0.224\textwidth]{Fig1/InSb_lnr_all.pdf}
         %\caption{(b)}
     %\end{subfigure}
     %\begin{subfigure}
     %    \centering
     %    \includegraphics[width=0.22\textwidth]{Fig1/InSb_phi_all.pdf}
         %\caption{(b)}
     %\end{subfigure}
     %\caption{Plots (a) and (b) show $|r|$ and $\varphi$ measurements for Si, respectively, at three relative positions: 0, 10, and 100 \textmu m. The known Si reflection coefficient of 0.55 is indicated by the dashed line in (a). Similarly, plots (c) and (d) display $|r|$ and $\varphi$ measurements for InSb at the same positions. The dashed line in (c) represents the reflectivity of InSb obtained from Drude model fitting.}
    % \label{fig:fig1}
%\end{figure}

%\colorbox{Green}{------------proofreading-------------}

A commercial THz instrument is used for both generating and detecting THz radiation using photoconductive antennas (TeraFlash Pro, TOPTICA Photonics Inc). A custom-made setup is built with a special sample stage which allows multiple degrees of freedom \cite{UDCM}. In particular, a piezoelectric stage with sub-micron precision was included to control the position of the sample in the direction perpendicular to the sample surface. Spectroscopy measurements were conducted on 500~\textmu m thick Si and InSb wafers, using a gold mirror as the reference. The optical properties of both samples have been previously reported in the literature. The results presented for Si refer to a specific case where the optical properties are constant over the measured frequency range. The selection of InSb aimed to assess the applicability of the method when dealing with a strong frequency-dependent response within the measurement range. Notably, InSb exhibits metallic behavior with a plasma frequency in the low THz range \cite{howells1996transient}. This material has been investigated thoroughly in recent years for the possibility of building terahertz meta-materials with tunable frequency response. For instance, the plasma frequency of InSb can be tuned by varying the temperature or doping of the sample.

THz-TDS requires the measurement of sample and reference waveforms. Typically, they are then converted into the frequency domain using fast Fourier transform algorithms. The reflection coefficient $\tilde{r}(\omega)=\widetilde{E}_\text{sample}(\omega)/\widetilde{E}_\text{reference}(\omega)$ contains information about the amplitude $|\tilde{r}(\omega)|$ and the phase $\varphi(\omega)$ of the electric field reflected from the sample. As mentioned previously, the sample and reference must be extremely well aligned for a reliable measurement. Despite meticulous efforts to ensure the utmost alignment between a sample and its reference, there remains some potential for a subtle shift in the sample position relative to the reference. The manual correction of this shift is inherently challenging as this directly affects the relative phase, without changing the amplitude. This behavior can be observed by measuring the samples at a starting position and then shifting them using the piezoelectric stage. The results for three different relative positions of 0, 10, and 100 \textmu m  between the sample and the reference are shown in Fig.~\ref{fig:fig1}. The left panels show the absolute value of the measured reflection coefficient $|\tilde{r}_m|$, while the right panels show the corresponding measured phase $\varphi_m$. While the absolute reflection is robust against misplacement, the phase is not. As seen from Fig.~\ref{fig:fig1}(b) and (d), the slope of the phase increases as the sample shifts away from the reference. This phase change results in a significant deviation from the true values while computing the optical properties. The deviation of the measured phase $\varphi_m$ from the intrinsic phase $\varphi_i$ of the sample (i.e., the phase corresponding to no misplacement between sample and reference) is given by
\begin{equation}\label{eq:phi_m}
\varphi_m(\omega)=\varphi_i(\omega)+\frac{\omega}{c}\frac{2l}{\cos{\theta}},
\end{equation}
%\cite{Zhang_2008}.
where $c$ is the speed of light in vacuum, $l$ is the sample-to-reference misalignment along the sample thickness, and $\theta$ is the angle of incidence of the THz pulse. The second term on the right-hand side of Eq.~\eqref{eq:phi_m} clearly shows why even a small misplacement $l\sim1$ \textmu m is problematic in these measurements: The $\omega/c$ coefficient is of the order of 20 rad/mm at 1 THz. Hence, at this frequency, the phase information is completely lost for a $\sim100$ \textmu m misalignment, and severely affected even by a $\sim10$ \textmu m misalignment if a broad THz range is of interest, like in our case. %Thus, in order to be able to successfully extract information from a THz spectroscopy measurement in reflection geometry, one must find a way to estimate the extrinsic contribution to the phase in order to retrieve the intrinsic sample phase $\varphi_i$.  

To address this issue, we consider the Kramers-Kronig relations, which are frequently utilized in optics to determine the complex values of physical parameters. In particular, these relations, which adhere to the principle of causality, establish a connection between the real and imaginary parts of the response function. Here we use their forms linking the magnitude of the reflection coefficient and its phase by representing the complex reflection coefficient $\tilde{r}=|\tilde{r}|e^{-i\varphi}$ in the logarithmic form as $\ln|\tilde{r}|-i\varphi$. This approach has been proposed and long discussed in the past decades \cite{smith1977dispersion,lovell1974applicationa, roessler1965kramerskronig, nash1995kramerskronig, peiponen1997ii, bertie1996accurate}. The two relations are
\begin{equation}\label{eq:kkr}
   \varphi(\omega) = \varphi_0 +  \frac{2\omega}{\pi}\textbf{P}\int_0^\infty\frac{ \text{ln}|\tilde{r}(\Omega)|}{\Omega^2-\omega^2}d\Omega,
\end{equation}
\begin{multline}
    \label{eq:ikkr}
  \ln\left|\tilde{r}(\omega)\right| - \ln\left|\tilde{r}(\omega^{\prime})\right| =\\ %\ln\left|\frac{\tilde{r}(\omega)}{\tilde{r}\left(\omega^{\prime}\right)}\right| =\\
  \frac{2}{\pi} \textbf{P} \int_0^{\infty} \Omega \varphi(\Omega)\left(\frac{1}{\Omega^2-\omega^2}-\frac{1}{\Omega^2-\omega^{\prime 2}}\right) d \Omega,
\end{multline}

From now on, we will refer to Eq.~\eqref{eq:kkr} as the \emph{direct} Kramers-Kronig relation and to Eq.~\eqref{eq:ikkr} as the \emph{inverse} Kramers-Kronig relation. In Eq.~\eqref{eq:ikkr}, $\ln|\tilde{r}(\omega')|$ corresponds to a reference reflection coefficient amplitude at any arbitrary frequency $\omega'$ \cite{UDCM, smith1977dispersion, peiponen1997ii, nash1995kramerskronig}. Theoretically, one could use the direct relation to solve the problem. However, as detailed in the companion paper \cite{UDCM}, we show that this method is less robust in extracting the correct phase than the inverse Kramers-Kronig relation. In the following, we therefore focus on the inverse method to correct the measured phase according to Eq.~\eqref{eq:phi_m}. % We consider in the following $\varphi_0$ = 0 rad, as this assumption is often verified in most materials \cite{smith1977dispersion, jepsen2019phase,mori2014progress,howells1996transient}. %since the phase must always be null at the DC limit \cite{jepsen2019phase}.

%As for $\ln|\tilde{r}_0|$, that is a constant factor that cannot be uniquely retrieved by the relations above, as discussed in detail in Ref.\cite{smith1977dispersion, peiponen1997ii, nash1995kramerskronig}. In our method, the knowledge of such a factor is not necessary, as we discuss below and in the companion paper \cite{UDCM}.
%As for $\ln|\tilde{r}_0(\omega)|$, it can be retrieved using any reliable data point from the measured $\tilde{r}$ at a given frequency $\omega_j$ using the relation 
%\begin{equation}\label{eq:ikkr_dev}
%\ln|\tilde{r}_0| = - \frac{2}{\pi}\textbf{P}\int_0^{\omega_\text{end}}\frac{\omega'\cdot\varphi_m(\omega')}{\omega'^2-\omega_j^2}d\omega' + \ln|\tilde{r}(\omega_j)|.
%\end{equation}
As shown in Fig.~\ref{fig:fig1}, $|\tilde{r}_{m}(\omega)|$ remains constant when the sample position $l$ is varied. 
In terms of the inverse Kramers-Kronig relation, the measured $\ln|\tilde{r}_{m}(\omega)|$ can be expressed as
\begin{multline}\label{eq:r_m}
    \ln \left|\tilde{r}_m(\omega)\right| = \ln \left|\tilde{r}_m(\omega^{\prime})\right| \\
    +\frac{2}{\pi} \textbf{P} \int_0^{\infty} \Omega \varphi_m(\Omega)\left(\frac{1}{\Omega^2-\omega^2}-\frac{1}{\Omega^2-\omega^{\prime 2}}\right)d \Omega.
\end{multline}
The absolute values of the measured reflection coefficient at $\omega$ and at $\omega^{\prime}$ are directly obtained from the measurement. On the right-hand side of the equation, the measured phase $\varphi_m$ contains the intrinsic phase $\varphi_i$ and $l$, where both terms are unknown and therefore inseparable without the knowledge of one or the other. It can be shown that the equality only holds when the integration is ideally performed up to infinity, thus making the effect of the misplacement negligible as shown in Appendix B in the companion paper \cite{UDCM}.

\begin{figure}[t!]
\centering
\includegraphics[width=\columnwidth]{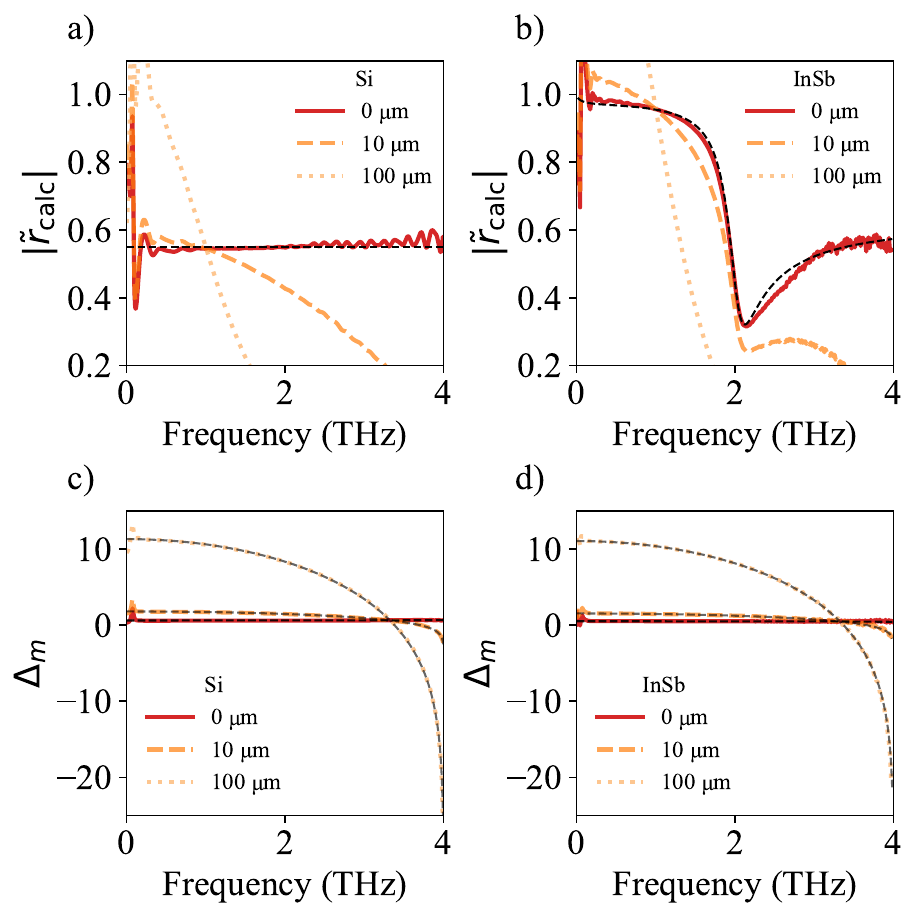}
\caption{Calculated reflection coefficients for (a) Si and (b) InSb using Eq.~\eqref{eq:r_calc} for three shifts of the sample position along the propagation direction. The dashed black line here is a fit to the data with no nominal shift. Calculated $\Delta_m$ from Eq.~\eqref{eq:dev} for (c) Si and (d) InSb, shown along with the fit for each plot, using the procedure discussed in the main text.}
\label{fig:rcalc_deltam}
\end{figure}

However, in any real measurement, the available range is restricted to a certain finite value.
%Hence, reasonable assumptions need to be formulated for the value of the integrals in the range from $\omega_\text{end}$ to $\infty$ to address this limitation.
%For the third term in Eq. \eqref{eq:r_m}, we can assume that the intrinsic phase $\varphi_i$ goes to zero away from resonances, as seen in Fig.\ref{fig:fig1}(b) and (d), and so does the integral as well. In other words, as long as the measurement range is sufficiently large to fully contain the response around possible resonances, there is no significant contribution from the intrinsic phase beyond $\omega_\text{end}$.
%On the contrary, the role of misplacement in the measured phase $\varphi_m$ is a linear function of $l$ and $\omega$.
%Since the Kramers-Kronig relations work on the principle of causality, Eq.\eqref{eq:r_m} would successfully provide the correct reflection coefficient if the measured phase was available over a very large frequency range, regardless of the misplacement as proved in the companion paper \cite{UDCM}.
In this case, the $l$-dependent term plays a significant role in determining the reflection coefficient, resulting in a poor reconstruction that does not match the expected curve. We choose to utilize this particular behavior of the inverse Kramers-Kronig relation to isolate and eliminate the value of $l$. For this purpose, we introduce a modified form of Eq.~\eqref{eq:r_m}, where we define a calculated reflection coefficient amplitude $\left|\tilde{r}_\text{calc}(\omega)\right|$ corresponding to the finite frequency range of the measurement:
\begin{multline}\label{eq:r_calc}
  \ln\left|\tilde{r}_\text{calc}(\omega)\right| \equiv
  \ln\left|\tilde{r}_m(\omega^{\prime})\right| \\+\frac{2}{\pi} \textbf{P} \int_0^{\omega_\text{end}} \Omega \varphi_m(\Omega)\left(\frac{1}{\Omega^2-\omega^2}-\frac{1}{\Omega^2-\omega^{\prime 2}}\right) d \Omega.
\end{multline}
Here, $\omega_\text{end}$ denotes the end of the measured frequency range which in our case is taken to be 4 THz. We choose $\omega'=1$ THz for the purpose of these measurements, as it is approximately the frequency with largest signal-to-noise ratio of our THz source. In principle however, any frequency where the data is accurately measured can be used.
%they cannot be expected to provide the correct reflection coefficient when the measured phase includes an $l-$ dependent term, which is purely an experimental artifact. However, 
%begin{align}\label{eq:rkkr_0_inf}
%\ln \left|\tilde{r}_{\mathrm{KKR}}(\omega)\right|&=  \ln \left|\tilde{r}_{0,\text{KKR}}\right|+\frac{2}{\pi} \mathbf{P} \int_0^{\omega_{\text{end}}} \frac{\Omega \cdot \varphi_i\left(\Omega\right)}{\Omega^2-\omega^2} d \Omega \nonumber\\
%&+\frac{4 l}{\pi c \cos \theta} \mathbf{P} \int_0^{\omega_{\text{end}}} \frac{\Omega^2}{\Omega^2-\omega^2} d \Omega,
%\end{align}
%exploiting the linearity of the inverse KKR operator. 
The computed $\left|\tilde{r}_{\mathrm{calc}}(\omega)\right|$ is now sensitive to the sample misplacement $l$, as shown in Fig.~\ref{fig:rcalc_deltam}(a) and (b) for Si and InSb, respectively. The value of $|\tilde{r}_\text{calc}(\omega)|$ using $\varphi_m$ measured at a finite displacement shows a significant deviation from the reference curve, indicated by the black dashed line. In Fig.~\ref{fig:rcalc_deltam}(a) such a line denotes the value of the reflection coefficient $|\tilde{r}|=0.55$ for Si as reported in the literature \cite{dai2004terahertz,ronne1997investigation}. In Fig.~\ref{fig:rcalc_deltam}(b), the black dashed line is obtained from fitting the data at $l=0$ to the Drude model for the measured reflectivity. The extracted parameters are $\varepsilon_{\infty} = 18.16$, $\omega_{p}/2 \pi = 2.005$ THz, and $\gamma/2 \pi = 0.26$ THz, consistent with the values reported in Ref.~\cite{howells1996transient,houver20192d}.
To isolate the value of $l$ from the measurement, it is necessary to compare these results with the experimental data for the reflection coefficient amplitude. The advantage of comparing $|\tilde{r}_m|$ and $|\tilde{r}_\text{calc}|$ is that the \emph{measured} reflection coefficient is primarily unaffected by the small shifts, whereas the \emph{calculated} $|\tilde{r}_\text{calc}|$ is highly dependent on it. 

%\begin{figure}[t!]
%\centering
%\includegraphics[width=\columnwidth]{PRL_delta_m.pdf}
%\caption{Calculated $\Delta_m$ from Eq.~\eqref{eq:dev} for (a) Si and (b) %InSb, shown along with the fit for each plot, using the procedure %discussed in the main text.}
%\label{fig:deltam}
%\end{figure}
%Noting that $\ln \left|\tilde{r}_{\mathrm{m}}(\omega)\right|$ corresponds to the first two terms in Eq. \eqref{eq:r_m}, where we assume the contribution of the third term to be negligible. Since the misplacement does not affect the modulus of the reflectivity, we write%
%\begin{multline}\label{eq:dev}
%\ln\left|\frac{\tilde{r}_\text{KKR}(\omega)}{\tilde{r}_{m}(\omega)}\right|=
%%\ln \left|\tilde{r}_{0,\text{KKR}}\right|-\ln \left|\tilde{r}_{0,m}\right|
%%\ln\left|\frac{\tilde{r}_\text{0,KKR}(\omega)}{\tilde{r}_{0,m}(\omega)}\right|
%\Delta_0+\frac{4l}{\pi c \cos \theta} \mathbf{P} \int_0^{\omega_\text{end}} \frac{\Omega^2}{\Omega^2-\omega^2} d \Omega  \\
%\approx \Delta_0+\frac{4l}{\pi c \cos\theta}\left[\omega_\text{end}+\frac {\omega}{2} \ln \left|\frac{\omega_{\mathrm{end}}-%\omega}{\omega_{\mathrm{end}}+\omega}\right|\right],
%\end{multline}

We can now use the discrepancy in the measured and calculated reflectivities $\left|\tilde{r}_{m}(\omega)\right|$ and $ \left|\tilde{r}_{\text{calc}}(\omega)\right|$ to estimate the misplacement $l$. The quantity to be computed is the difference between $\ln \left|\tilde{r}_{\text{calc}}(\omega)\right|$ and $\ln \left|\tilde{r}_{m}(\omega)\right|$:
\begin{multline}
    \label{eq:r_calc_diff}
  \ln\left|\tilde{r}_\text{calc}(\omega)\right| - \ln\left|\tilde{r}_m(\omega)\right| =
  -\ln\left|\frac{\tilde{r}_m(\omega)}{\tilde{r}_m(\omega')}\right| \\
  +\frac{2}{\pi} \textbf{P} \int_0^{\omega_\text{end}} \Omega \varphi_m(\Omega)\left(\frac{1}{\Omega^2-\omega^2}-\frac{1}{\Omega^2-\omega^{\prime 2}}\right) d \Omega \approx \\
  \frac{2l}{\pi c \cos\theta}\left(\omega \ln \left|\frac{\omega_{\mathrm{end}}-\omega}{\omega_{\mathrm{end}}+\omega}\right|- \omega' \ln \left|\frac{\omega_{\mathrm{end}}-\omega'}{\omega_{\mathrm{end}}+\omega'}\right|\right) + C,
\end{multline}
where the approximate analytical solution has been obtained thanks to Eq.~\eqref{eq:phi_m}, and $C$ represents the integration error introduced for reducing the upper limit of the integration range to $\omega_{\mathrm{end}}$ \cite{UDCM}. There are two ways of finding $l$: experimentally, by scanning the sample position over a range of $l$ values until a minimum is found, or analytically. Both methods are described in detail in the companion paper \cite{UDCM}. Here, we report the main results of the analytical approach. Since the integral shown in Eq.~\eqref{eq:r_calc_diff} has a finite frequency range, the terms inside the bracket can be split into two separate integrals, one depending on $\omega$, and the other on $\omega'$. The terms containing an $\omega'$ dependence can be grouped together and are described by the constant $C'$.  Using the remaining terms, we define $\Delta_m$ as
\begin{multline}\label{eq:dev}
    \Delta_m\left(\omega\right) \equiv
    \frac{2}{\pi} \textbf{P} \int_0^{\omega_\text{end}} \frac{\Omega \varphi_m(\Omega)}{\Omega^2-\omega^2} d \Omega - \ln\left|\tilde{r}_m(\omega)\right| \approx\\
    \frac{2l\omega}{\pi c \cos\theta} \ln \left|\frac{\omega_{\mathrm{end}}-\omega}{\omega_{\mathrm{end}}+\omega}\right| + C + C'.
\end{multline}

This result can be used to extract the value of $l$. The calculated $\Delta_m$ is fitted using the analytical formula in the second line of Eq.~\eqref{eq:dev}, where $l$ and $C+C'$ are used as free parameters.
%The third line of Eq.~\eqref{eq:dev} can be used as the fitting function for $\ln\left|{\tilde{r}_\text{KKR}(\omega)}/{\tilde{r}_{m}(\omega)}\right|$, with $\Delta_0$ and $l$ as the free fitting parameters.
The fitted curves for $\Delta_m$ for Si and InSb are shown in Fig.~\ref{fig:rcalc_deltam}(c) and (d) as black dashed lines, respectively. The fit is performed over the entire frequency range for the Si measurement, while for InSb, the range 0.25 -- 3.5 THz is used since. For Si, and for the 0, 10, and 100 \textmu m positional shift of the stage, the estimated misplacement values $l$ are $-0.14$ \textmu m, $+11.21$ \textmu m and $+101.18$ \textmu m, and the extracted values of $C+C'$ are 0.59, 1.78 and 11.27, respectively. For InSb, and the same nominal shifts, $l$ is estimated to be 0.29 \textmu m, 10.1 \textmu m, and 100.06 \textmu m, while $C+C'$ are found to be 0.49, 1.52, 11.02. For both samples, the estimated misplacement is robust and close to the nominal value. Using these values of $l$ in Eq.~\eqref{eq:phi_m}, the corrected intrinsic phase $\varphi_i$ can be retrieved from $\varphi_m$.

\begin{figure}[t]
\centering
\includegraphics[width=\columnwidth]{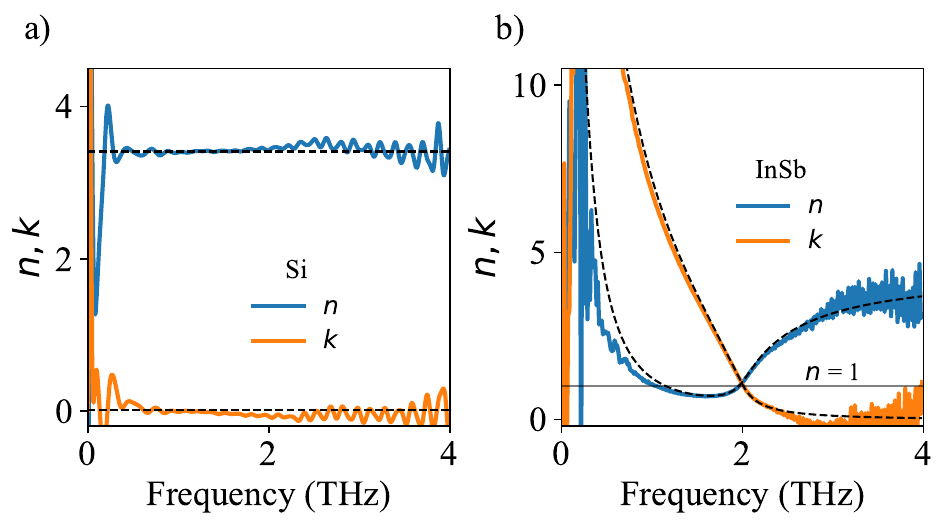}
\caption{Complex refractive index obtained from the corrected phase for (a) Si and (b) InSb. In both panels, blue and orange solid lines denote the real part $n$ and, respectively, the imaginary part $k$ of the complex refractive index. Black dashed lines in (a) are based on literature values. In (b), the black dashed lines are from the fit of the data to the Drude model.}
\label{fig:nk}
\end{figure}

Now, $|\tilde{r}_m|$ and $\varphi_i$ can be used to extract the complex refractive index of the material using the relations
\begin{subequations}
\label{eq:refr_index}
\begin{eqnarray}
n&=&\frac{1-|r|^{2}}{1+|r|^{2}-2|r|\cos\varphi}, \label{eq:refr_index_n}\\
k&=&\frac{2|r|\sin \varphi}{1+|r|^{2}-2|r|\cos\varphi}, \label{eq:refr_index_k}
\end{eqnarray}
\end{subequations}
derived from the Fresnel equations for the case of normal incidence. The computed real and imaginary parts of the complex refractive index are plotted in Fig.~\ref{fig:nk}. The results for Si in Fig.~\ref{fig:nk}(a) show a good agreement with the previously reported values in literature indicated by the black dashed line \cite{dai2004terahertz,ronne1997investigation}. In order to estimate the accuracy of our method, we averaged the extracted refractive index over a chosen frequency range and calculated the relative error with respect to literature values. For silicon, $n_{\textrm Si}\approx3.42$ is known to be to a good approximation in this frequency range \cite{dai2004terahertz}. We find the value $n_{\textrm Si, exp}\approx3.33$, i.e., with an accuracy better than 2.5\% in the 0.25--3.5 THz range. By considering only the central frequency range 1--2 THz, where the signal-to-noise ratio of our instrument is highest, we obtain $n_{\textrm Si, exp}\approx3.39$, with an accuracy better than 0.6\%.

The data also show an accurate extraction of the response of InSb over a broad frequency range, with a clear signature of the plasma edge at around 2 THz, in agreement with the prediction from the Drude model. Furthermore, in the region from 1.05 THz to 1.95 THz, the real part of the refractive index is found to be less than one, with a minimum value of 0.691 at 1.58 THz. This is a manifestation of near-zero index (NZI) behavior. These materials have unique optical properties, where their low refractive index can lead to exotic optical effects \cite{wu2021epsilonnearzero}. While implicit in the response expected within the Drude model \cite{houver20192d, chang2012complex}, to the best of our knowledge, direct experimental evidence of this remarkable condition has not been reported yet for this material.

\begin{figure}[t!]
\centering
\includegraphics[width=\columnwidth]{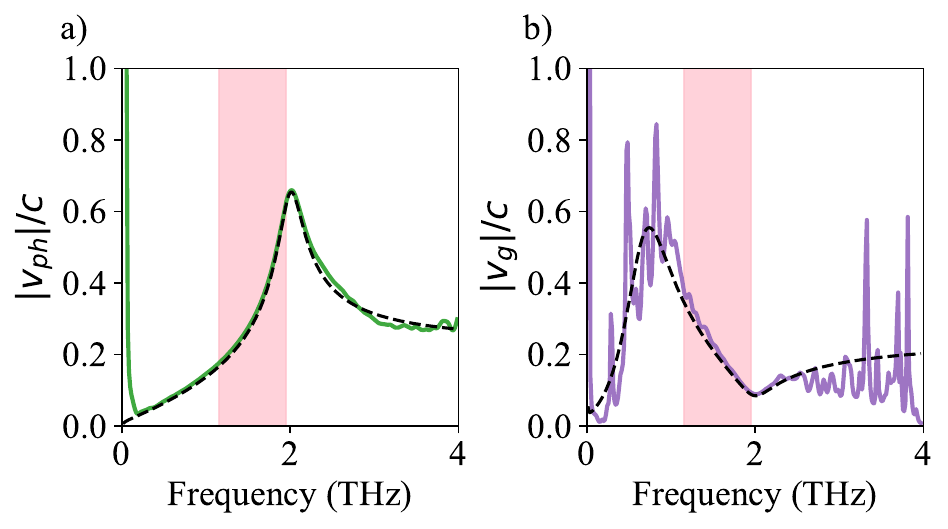}
\caption{(a) Phase velocity (green solid line) and (b) group velocity (purple solid line) for InSb. Black dashed line: calculated values from the Drude model. The pink-shaded area denotes the NZI region.}
\label{fig:vgvp}
\end{figure}
To further investigate the consequences of NZI behavior, we calculate the phase and group velocities in the material using the relations $\tilde{v}_{p}(\omega) =c/\tilde{n}$ and $\tilde{v}_{g}(\omega)= {c}/({\tilde{n}+\omega\,{d\tilde{n}}/{d\omega}})$, respectively \cite{liberal2017nearzero}. To obtain a reliable estimate of the group and phase velocity for InSb, a low-pass filter was applied to the complex refractive index data to eliminate high-frequency oscillations caused by the differential term in the group velocity expression. The absolute values of the velocities obtained are plotted in Fig.~\ref{fig:vgvp}, normalized to the velocity of light $c$. The shaded region indicates where the refractive index is less than 1. In a lossless medium ($k=0$), the phase velocity would diverge as $n\rightarrow0$. The group velocity on the other hand would tend to 0 \cite{liberal2017nearzero}. In our case, InSb has significant losses in the THz range, which dampens the divergence of the velocities.
%Based on the two equations, one can infer that for an ideal NZI material ($k=0$), as $\omega \rightarrow \omega_{p}$, the refractive index $n(\omega) \rightarrow 0$. This makes the phase velocity diverge. However, the derivative of the slope could be non-zero. This results in the group velocity $v_{g} \rightarrow 0$ as $n \rightarrow 0$. For the case of InSb, since $k$ is large,

The divergence of the phase velocity results in a maximum amplitude of $0.66c$ in correspondence of the plasma frequency. At the same frequency, the group velocity reaches a value as low as $0.08c$ at 2 THz, which could be of interest for observing slow light or trapping effects \cite{kinsey2019nearzeroindex}. Achieving slow light often requires the fabrication of structured materials. However, it is remarkable that the value observed here is approximately half of the minimum group velocity of indium tin oxide (ITO) films, which are known for their NZI behavior \cite{vertchenko2023nearzeroindex, alam2016large}. The crossing point of $n$ and $k$, where Re($\varepsilon$)=0, also occurs close to $\omega_p$ where $n=k=1.08$ at $1.99$ THz, as seen in Fig.~\ref{fig:nk}. It is important to note that the properties of InSb are highly tunable based on factors such as temperature or doping concentration, as previously reported \cite{howells1996transient, chochol2016magnetooptical, law2014doped}. This would enable the utilization of InSb for making metamaterials as well as for NZI applications, such as beam steering and field enhancement and observing superradiance.

To summarize, we demonstrated a phase-correction technique based on the inverse Kramers-Kronig relations. This method uses non-iterative calculations on a single THz time-domain signal with arbitrary misplacement and allows to retrieve accurate optical properties of the measured material with no need of heavy computational methods. We tested the method on both a sample with constant refractive index and negligible losses, as well as with a material with a strong resonance in the measurement range. In the case of InSb, we report direct experimental observation of near-zero index behavior in the THz range. Our method opens up for widespread use of THz reflection spectroscopy in all materials beyond those that are transparent to this frequency range. Of particular interest are quantum materials which host a plethora of collective excitations at terahertz frequencies, often linked to their exotic properties. 

We thank R. Knut and H. Rostami for the useful discussion. T.-T. Yeh and S. Bonetti acknowledge support from the Knut and Alice Wallenberg Foundation, grant No. 2019.0068. M. Pancaldi and S. Bonetti acknowledge support from the Italian Ministry of University and Research, PRIN2020 funding program, Grant No. 2020PY8KTC.
\providecommand{\noopsort}[1]{}\providecommand{\singleletter}[1]{#1}%

\end{document}